\documentclass[reprint,onecolumn,aps,prl,superscriptaddress,nobibnotes,amssymb,floatfix]{revtex4-1}

\usepackage{gensymb}
\usepackage{graphicx}
\usepackage[colorlinks=true,allcolors=blue]{hyperref}
\begin{document}

%\preprint{}

\title{Re-enterant efficiency of phototaxis in Chlamydomonas reinhardtii cells}


\author{Sujeet Kumar Choudhary}
\affiliation{Department of Physics, Indian Institute of Science, Bangalore, India}
\author{Aparna Baskaran}
\affiliation{Martin Fisher School of Physics, Brandeis University, USA}
\author{Prerna Sharma}
\affiliation{Department of Physics, Indian Institute of Science, Bangalore, India}
%\thanks{}


\date{\today}

\begin{abstract}
Phototaxis is one of the most fundamental stimulus-response behaviors in biology wherein motile micro-organisms sense light gradients to swim towards the light source. Apart from single cell survival and growth, it plays a major role at the global scale of aquatic ecosystem and bio-reactors. We study photoaxis of single celled algae Chalmydomonas reinhardtii as a function of cell number density and light stimulus using high spatio-temporal video microscopy. Surprisingly, the phototactic efficiency has a minimum at a well-defined number density, for a given light gradient, above which the phototaxis behaviour of collection of cells can even exceed the performance obtainable from single isolated cells. We show that the origin of enhancement of performance above the critical concentration lies in the slowing down of the cells which enables them to sense light more effectively.
We also show that this steady state  phenomenology is well captured by a modelling the phototactic response as a density dependent torque acting on an active Brownian particle. 

%21 october to 25 nov
%Many  photosynthetic micro-organism respond to the light and move towards or away from the light gradient, a phenomena called phototaxis. It plays important role in aquatic ecosystem by affecting phytoplancton mass transfer through diel vertical migration and algal bloom, and has major applications in bioreactor, microbiopropellers and artificial microswimmers. Previous studies  show that phototactic response of \textit{Chlamydomonas reinhardtii} enhances  with cell density due to collective effect. As these results were drawn from population level measurement, phototactic response at very dilute cell suspension (up to single cell)  are missing. However phototactic response have been extrapolated to zero in low cell density limit.  Here using high spatio-temporal resolution data, we show that there exist a critical cell density ($\rho_c$) below which phototactic response diverges from zero, manifesting the enhancement in  response to the light stimulus at lower cell concentration. Our finding show that low density regime is governed by different phenomena other than collective behaviour. We then use steady state distribution of Active Brownian Particle (ABP) model under external torque to fit (or reproduce) the experimental data.    
\end{abstract}

%\keywords{}

\maketitle

\section{Introduction}

Collective behaviour  is observed in biological systems at different levels of biological organization from cells in tissues to colonies of microorganisms to flocks or herds of macroscopic animals \cite{Parrish2002,Sumpter,Kearns2010}. Phenomena at the level of the population in such systems cannot always be predicted by simply knowing the behaviour of individuals. For example, biofilms of Bacillus subtilis bacteria exhibit oscillatory growth rate whereas no such oscillations exist in dilute suspensions of the same bacteria \cite{bacillus1}. Collective behavior in microorganisms is of particular interest as it can be thought of as a precursor to multicellularity and more complex organizations of living systems. Consequently, a number of quantitative studies have recently elucidated the origin of collective phenomena in a wide variety of micro-organisms such as E-coli \cite{ecoli1,ecoli2}, Bacillus subtilis \cite{bacillus1,bacillus2,bacillus3}, Synechocystis sp. \cite{cyno1,cyno2}, Pseudomonas \cite{pseudonomas1,pseudonomas2} and Myxococcus xanthus \cite{myxo1,myxo2}.

Taxis, a transport phenomenon in which organisms undergo directed movement in response to a stimulus or a nutrient gradient, provides a particularly tractable context in which to explore collective behaviour. As a particular example, phototactic cells such as algae and cyanobacteria respond to light gradients  \cite {Feinlib1,Feinlib2}. Single celled eukaryotic algae Chlamydomonas reinhardtii (CR) is a model biological organism for studying phototaxis \cite{goldstein2015}. While single cell response of CR  to light can be tuned by varying physical variables such as light intensity, fluid viscosity as well as through chemical variables such as extracellular calcium concentration \cite{Feinlib2,Giometto7045,Viscosity_Volvox,Stavis367,dolle}. It was shown recently that phototaxis of dense suspensions of CR was governed by the cell number density itself revealing that collective effects could modulate the single cell response \cite{Furlan2012}. Here, we set up quasi-two-dimensional phototaxis assay with CR to study the cross-over from the individual to collective phototaxis and identify the mechanisms underlying the emergence of its collective phototaxis. 

CR has two flagella and an eye-spot located near the cell equator. Its flagella move in breast-stroke fashion to propel the cell body through the fluid \cite{Witman,goldstein2015}. The ellipsoidal shaped cell body rotates about its own axis while swimmng enabling the eyespot to scan the incident light around the swimming path \cite{Foster1980,Eyespot1,Jekely2008}. Under phototactic light exposure, beating of the flagellum closest to the eyespot is inhibited whereas beating of the further away one is enhanced resulting in aligning the cell towards the light source \cite{Ruffer,Howard}. We use a high speed camera to record individual trajectories of hundreds of cells under varying light intensities and cell concentrations. 

We find that starting from few cells per unit volume, phototatic efficiency decreases with increasing cell concentration until a critical concentration is reached above which the efficiency increases with increasing concentration. Thus, the phototactic efficiency is a reentrant function of the cell density. We further show that the origin of this reentrant behavior lies in the decrease in the swim speed of the cells as density increases beyond the critical concentration. Finally we find that the observed phenomenology is well captured by a model of active Brownian particles subject to a density dependent external torque.

\section{Experimental details}

CC-1690 (wild type) cells were used for the experiment. Synchronous culture of CR were grown in TAP media at 25$\,^{\circ}\mathrm{C}$ on 12 h/12 h light/dark cycle in an orbital shaker (135 rpm). Fig. 1a shows schematic of experimental setup. Cell suspension was observed in rectangular quasi-two-dimensional chambers (50 mm $\times$ 5 mm $\times$ 66 $\mu m $ ) made of glass slide and cover slip with double sided tape as a spacer. A blue laser beam of wavelength 488 $nm$ from the optical fiber illuminated one end of the chamber to act as a stimulus for phototaxis. Cell trajectories were imaged using bright field imaging with red light (760 $nm$ and above) illumination set up on an Olympus IX73 inverted microscope. Images were recorded at 100 frames per second at $\times$10  magnification using PCO 1200hs CMOS camera coupled to the microscope. $\times$10  objective has a large depth of focus that enables us to capture 2-D projections of the cell trajectories for as long as typically $\sim$20 seconds. Particle tracking was performed using image processing code in MATLAB and Python. For a given cell concentration and light intensity, 500-2500 trajectories were analysed to have robust statistics. 

%Cells were left 1 h under dark condition before being used in the experiment. %Cells were used for experiment between 48 h and 96 h after inoculation in liquid medium to ensure reproducibility  

\section{Results}

Cells move in random directions in the absence of blue light (Fig. 1b). Presence of blue light at one end of the chamber biases the movement of a majority of cells towards the light source (Fig. 1c). However, a small but finite fraction of cells continue to move in directions other than the direction of light source (Fig. 1c). Probability density  as a function of polar angle in the plane characterizes this phenomenon quantitatively (Fig. 1d). The distribution is, naturally, peaked in the source direction with the peak height increasing with increasing light intensity (Fig. 1d). In order to analyze the response of the system  tractably, we define phototactic efficiency, $\zeta$, as the fraction of cells that move in a direction $\pm 15$ degrees of the source direction. At low intensities of the light source, $\zeta$ is significantly less than 1 and approaches 1 at higher intensities (Fig. 1d inset).

While the phototactic efficiency shows the anticipated increase with increasing light intensity, one expects that cell concentration will also play a role in governing phototatxis at the population level \cite{Furlan2012}. Fig. 2 a-d show representative cell trajectories as a function of cell number density for a fixed light intensity. Starting from suspensions of few cells, the peak height of the probability density decreases with increasing concentration until a critical concentration $\rho_c$ is reached. Above $\rho_c$, the peak height increases monotonically with the concentration (Fig. 2e). This re-entrant phototaxis behaviour can equivalently be represented by the non monotonic variation of $\zeta$ with cell concentration (Fig. 2f).

It could be reasonably expected that the measured probability distribution of trajectory orientations $\psi(\theta)$ could be captured by a self-propelled particle model \cite{PhysRevE.92.052143}. The simplest such model in this context would be that of non-interacting active Brownian particles subject to a polar aligning torque that tends to turn the trajectories of the particles along some particular direction in the lab frame. Let us pick this direction to be along $\theta=0$. The Fokker-Planck equation governing the dynamics of the probability density $\psi(\theta,t)$ for the orientations of these self-propelled particles is given by,
\begin{equation}
\partial_t \psi(\theta,t) = D_R {\partial_\theta}^2 \psi+\frac{\gamma}{\xi_r} \partial_\theta(sin\theta\psi)
\end{equation}
where $D_R$ is the rotational diffusion coefficient, $\gamma$ is the torque strength and $\xi_r$ is the rotational friction coefficient. The steady state solution to this equation is the well known Von-Mises distribution function of the form, $\psi(\theta) = \frac{e^{\kappa cos\theta}}{2\pi I_0(\kappa)}$ where $\kappa = \frac{\gamma}{D_R}$ and $I_0$ is  modified Bessel's function of first kind. The experimentally obtained probability density as a function of polar angle is well fit by the Von-Mises distribution (Fig. 3a). The density dependence of this probability distribution can now arise either through $D_R$, implying that the rotational diffusion and hence the characteristic decorrelation time of the orientational autocorrelation function depends on density, or through the torque $\gamma$. The experimental data reveals that this decorrelation time is independent of cell concentration (Fig. 3a inset). Therefore one can extract an effective density dependent torque acting on the cells by fitting the experimental distribution to the Von-Mises distribution.  The variation of best-fit values of $\gamma$  with cell concentration (Fig. 3b) is qualitatively similar to that of the previously shown model independent phototactic efficiency $\zeta$. Therefore, the reentrant behavior of the phototactic efficiency as a function of density is reliably captured by modelling this collective phenomenon as an effective density dependent torque on each cell.

While one could potentially rationalize the decrease in phototactic efficiency as the concentration increases as an effect of cell-cell scattering, the increase in $\zeta$ at densities greater than the critical concentration is more puzzling. It may be reasonable to postulate that at high densities its primary effect on the behavior of a single cell is that it slows down and indeed that is the case in our experiments (Fig. 4a). This has been referred to as density dependent motility in the context of the active matter literature \cite{Cates_ar,Cates11715} . This could potentially affect the phototactic efficiency because of  how CR cells detect light. The cells follows a helical trajectory due to cell body rotation. The cell body rotation allows the cell to collect photons from all directions in space. A decrease
in linear speed implies a decrease in cell body rotation rate which enables the cell to collect more photons per unit time and therefore detect the light direction more
accurately (Fig. 4d).

One way to possibly validate this postulated mechanism for the increase in $\zeta$ as  density increases beyond $\rho_c$ would be to slow the cells down without changing the concentration of cells. One of the simplest ways to achieve that is to add polymer to the suspension medium which increases the drag force on the cells, thereby lowering their speed. We use varying concentrations of methylcellulose to tune the speed of the cells keeping cell concentration fixed (Fig. 4b). We find that $\zeta$ increases with increase in methycellulose concentration, confirming the hypothesis that the observed increase in $\zeta$ with increasing cell concentration is mainly due to lowering of cell speed (Fig. 4c).

\section*{Discussion}

To summarize, we find that phototactic efficiency of CR cells is re-entrant in going from low density dilute regime to high density collective one wherein dilute suspensions have smaller efficiency than that of single cell limit and dense suspensions have the opposite trend. We have identified the mechanism of enhanced efficiency in the collective regime to be the decrease in linear speed of the cells as the concentration increases. We speculate that decrease in linear speed leads to a decrease in rotational speed of the cells that enables them to sense the light direction more accurately.

The cell speed is nearly independent of concentration below the threshold concentration that marks the crossover between the individual and collective behavior. Therefore, the mechanism for decrease in efficiency with increasing cell concentration in the dilute regime is likely to be some other form of hydrodynamics interaction or steric in nature. It also remains to be explored how tightly the single cell response is coupled with its collective response. In other words, how chemical or genetic modifications that alter the single cell phototaxis efficiency affect the collective behavior of such modified cells.

 Complexity is common in biological systems and often its origin is difficult to identify. Our results have demonstrated a rather simple physical and phenomenological mechanism underlying the observed complexity in the collective phototaxis of CR cells. Apart from identifying a particular phenomenology associated with zooplanktons with a single eye spot, this work can serve as a paradigm for analysis of collective motility and taxis in microorganisms in general and perhaps motivate design of control algorithms in collective robotics.\\

\textbf{Author Contributions}: AB and PS designed the research, SKC carried out the experiments and associated analysis. All authors wrote the article.

\textbf{Acknowledgement}
AB acknowledges support from Brandeis Center for Bioinspired Soft Materials, NSF MRSEC, DMR-1420382 and the hospitality of IISc and IMSc where part of this work was completed. This work was supported by the Wellcome Trust/DBT India Alliance Fellowship [grant number IA/I/16/1/502356] awarded to P. Sharma.

\bibliography{Final_Ref}

%\newpage   

  \begin{figure}
 \includegraphics[scale=0.55]{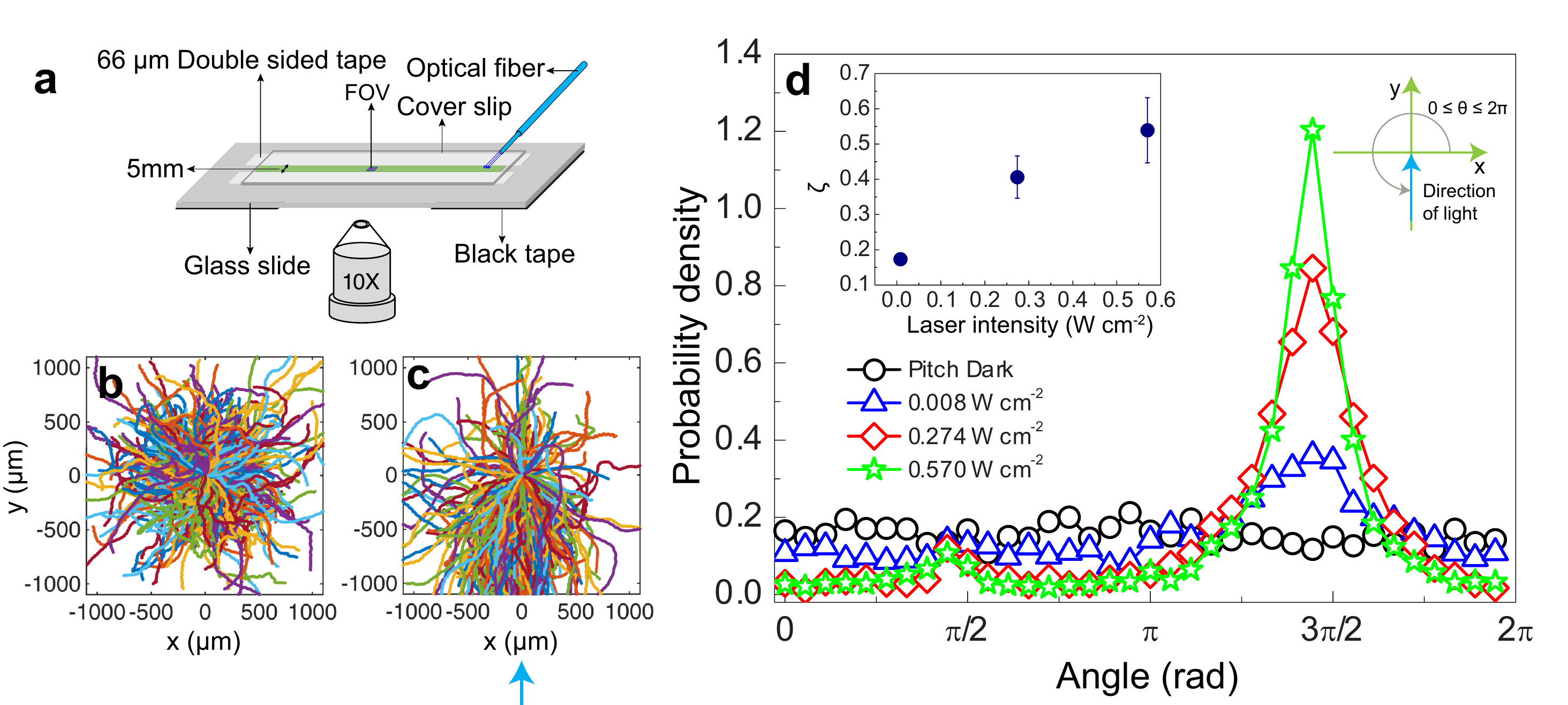}%
 \caption{\label{int_pdf} \textbf{Experimental setup and Phototactic response.} (\textbf{a}) Schematic of the experimental setup (\textbf{b}) Trajectories of \textit{Chlamydomanas reinhardtii} in pitch dark condition. Each trajectory is arbitrarily coloured for visual clarity. (\textbf{c}) Trajectories in presence of light ($I = 0.2737\, W\, cm^{-2}$). Blue arrow at the bottom  of Fig. 1(\textbf{c}) shows the direction of stimulus light. In the pitch dark condition, cell trajectories are uniform in all directions whereas in presence of light, a large fraction of cell trajectories are oriented towards the light source (positive phototaxis). (\textbf{d}) Probability density $P(\theta)$ for different light intensities according to the sign convention given at the top right corner in the Fig. 1(\textbf{d}). (Inset) Phototactic efficiency $\zeta$ as a function of light intensity.  The error bars correspond to the standard deviation of $\zeta$.}
 \end{figure}
 
  \begin{figure}
 \includegraphics[scale=0.6]{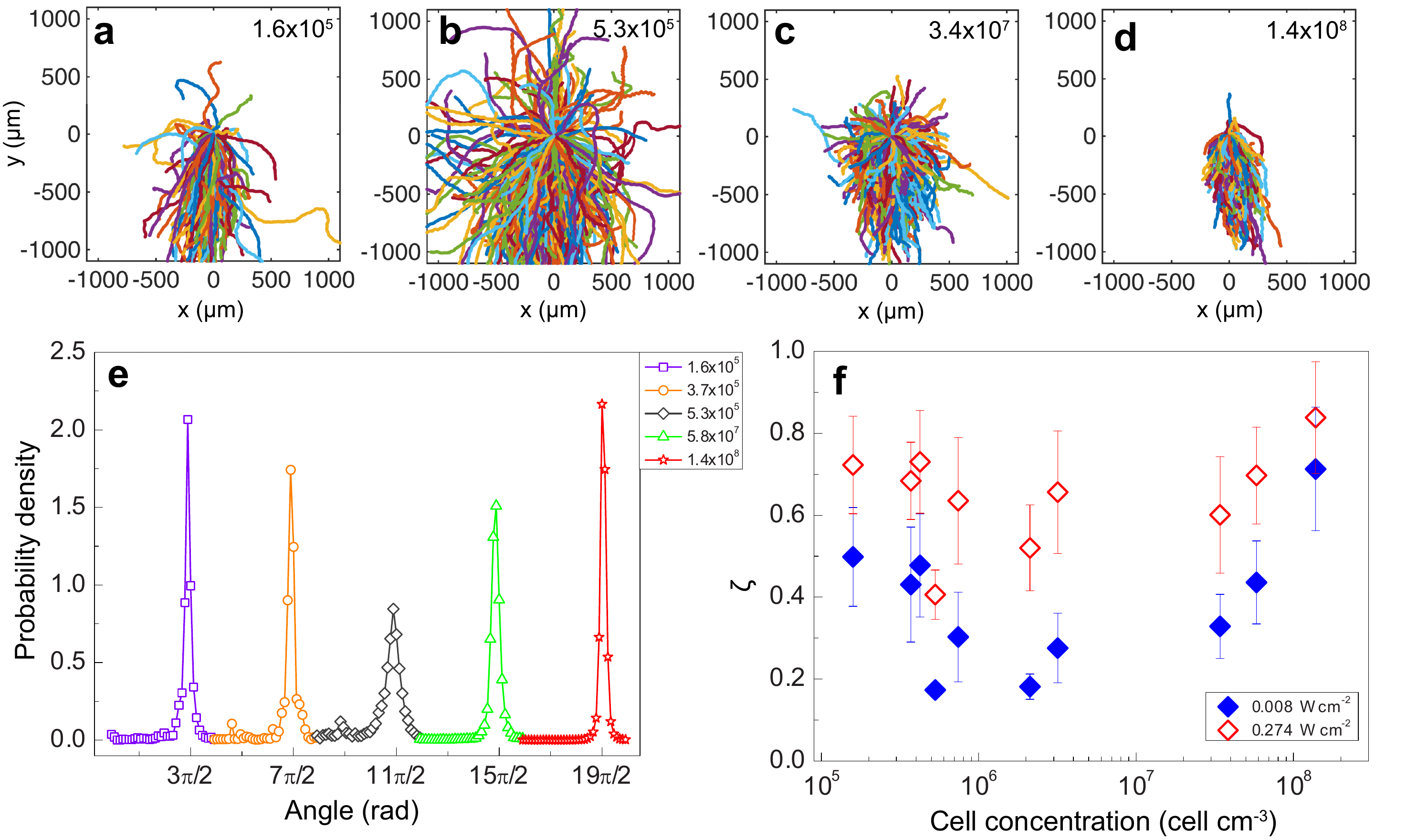}%
 \caption{\label{conc_pdf} \textbf{Phototactic efficiency  is reentrant with the cell concentration.} (\textbf{a-d}) Typical cell trajectories of \textit{Chlamydomanas reinhardtii} under fixed light intensity ($I = 0.2737\, W\, cm^{-2}$) with varying cell concentration (Legend in unit of $cells\,cm^{-3}$). (\textbf{e}) Probability density $P(\theta)$  under fixed light intensity ($I = 0.274\, W\, cm^{-2}$) for increasing cell concentration. The angle $\theta$ has been offset by multiple of $2\pi$ to shift the peak position for clarity. The height of the peaks quantify the reentrant behaviour of phototactic efficiency (Legend in unit of $cells\,cm^{-3}$). (\textbf{f}) Phototactic efficiency $\zeta$ as a function of cell concentration corresponding to two different light intensities. The error bars correspond to the standard deviation of $\zeta$. Phototactic efficiency decreases with increasing cell concentration until a critical concentration ($\rho_c$) reached, above which phototactic efficiency increases with cell concentration. The dependence of phototactic efficiency on cell concentration is stronger at the lower light intensities.}
 \end{figure}

   \begin{figure}
 \includegraphics[scale=0.6]{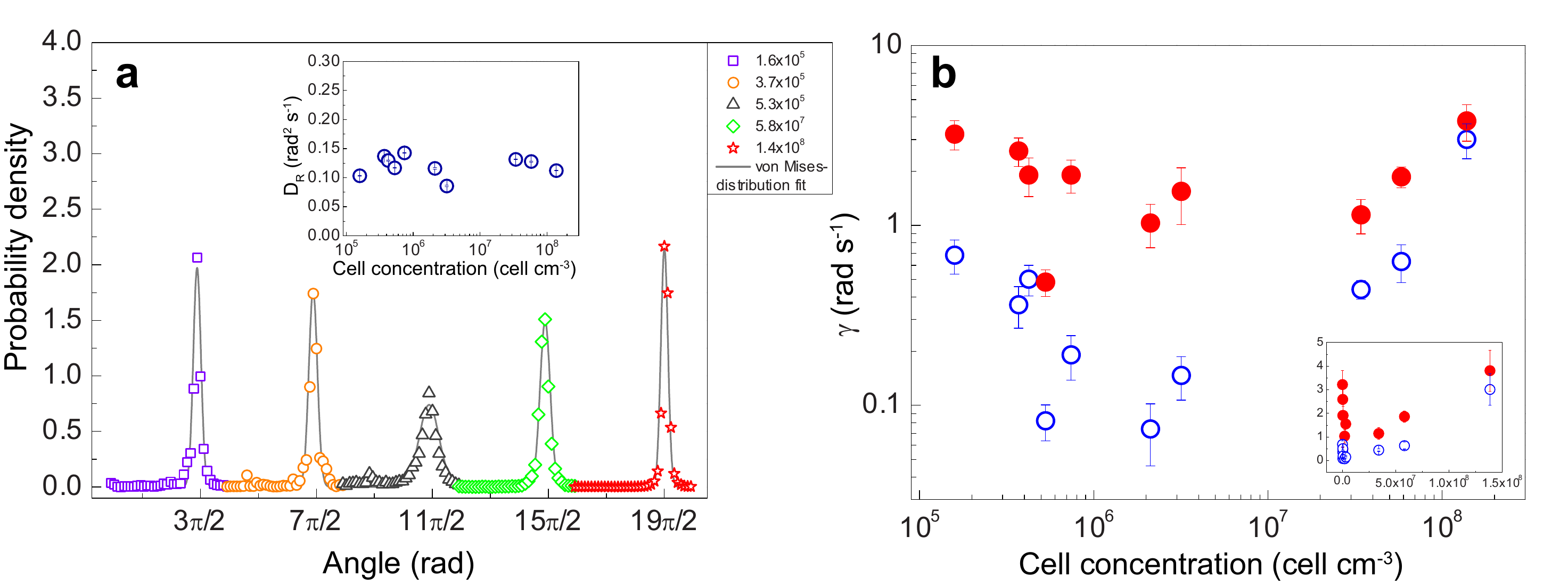}%
 \caption{\label{Dr_gamma} \textbf{Self propelled particle model for collective phototaxis} (\textbf{a}) Fit of  the experimental probability densities (Fig 2(\textbf{e})) to von Mises distribution, $\psi(\theta) = \frac{e^{\kappa cos\theta}}{2\pi I_0(\kappa)}$  where $\kappa = \frac{\gamma}{D_R}$  with rotational diffusion coefficient $D_R$, $0.13\, rad^2\,s^{-1}$. The angle $\theta$ has been offset by multiple of $2\pi$ to shift the peak position for clarity (Legend in units of $cell\,cm^{-3}$). (\textbf{Inset}) Plot of rotation diffusion coefficient, $D_R$ with the cell concentration shows that, $D_R$ is independent of the cell concentration. (\textbf{b}) A Log-log plot of torque strength  $\gamma$ as a function of cell concentration under two different light intensities $I_1 = 0.008\, W\, cm^{-2}$ (Open circle) and $I_2 = 0.274\, W\, cm^{-2}$ (Closed circle). The error bars correspond to the standard deviation of $\gamma$. The reentrant phototaxis behaviour observed in the experiment can be effectively captured by a density dependent aligning torque. (\textbf{Inset}) A linear representation of the same data.}
 \end{figure}
 
    \begin{figure}
 \includegraphics[scale=0.55]{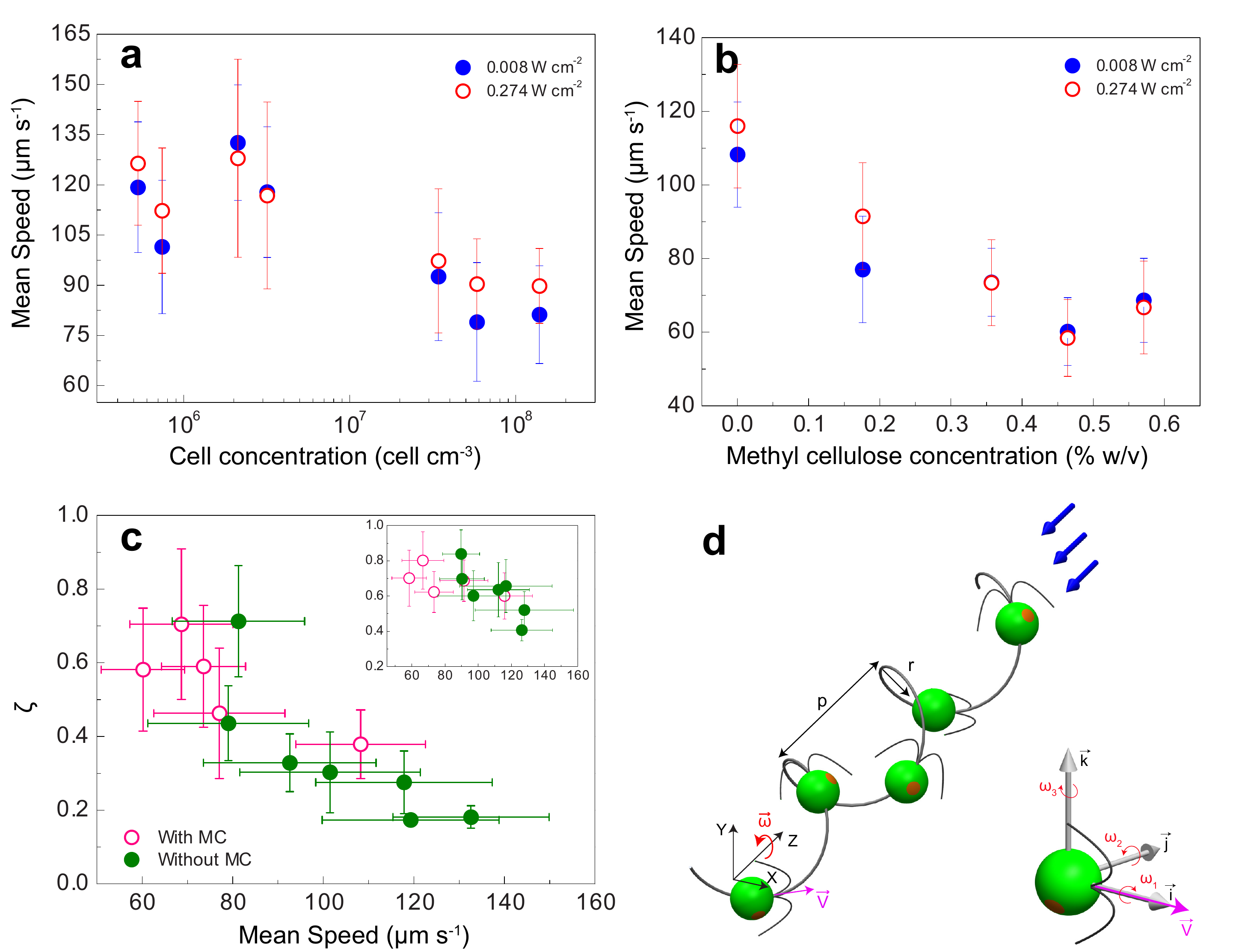}%
 \caption{\label{MC} \textbf{Physical origin of reentrant phototactic efficiency} (\textbf{a}) Mean speed as a function of cell concentration above critical concentration ($\rho_c$). As cell concentration increases,  cells slow down. (\textbf{b}) Mean speed as a function of Methyl cellulose concentration at a constant  cell concentration  $2.9\times10^6\, cell\,cm^{-3}$. Cell's speed was slowed down using Methyl cellulose in the suspension medium. (\textbf{c}) Phototactic efficiency $\zeta$ as a function of cell's mean speed at the intensity $I_1 = 0.008\, W\, cm^{-2}$ . Pink open circle (colour online)  correspond to the phototactic efficiency when cells are slowed down using methyl cellulose and green filled circle (colour online)   correspond to the phototactic efficiency when cell's speed was varied by cell concentration. In both the cases phototactic efficiency decreases as the cell's speed increases. Phototactic efficiency is controlled by mean speed. (\textbf{Inset}) Phototactic efficiency $\zeta$ as a function of cell's mean speed at the intensity $I_2 = 0.274\, W\, cm^{-2}$. (\textbf{d}) Schematic of a \textit{Chlamydomonas} cell trajectory illustrating $V = \frac{|\vec{\omega}|}{2\pi} p$. Slower cells turns slowly. (\textbf{Inset}) Definition of angular and linear velocity component along the body axes of the cell. Helical trajectory results whenever $\vec{\omega}$ is neither parallel nor perpendicular to $\vec{V}$ ($\omega_1\, \neq\,0\, , \sqrt{{\omega_2}^2+{\omega_3}^2}\, \neq\,0$). Error bars in all the plots correspond to the standard deviation of respective physical quantities.}
 \end{figure}

\end{document}